\documentclass[]{pasj01}

\newenvironment{trueauthors}{\section*{Author contributions}\fontsize{8}{11}\selectfont}{\par}

\begin{document} 
\Received{2017/06/04}
\Accepted{2017/06/30}
\jyear{2017}

\title{Search for Thermal X-ray Features from the Crab nebula with Hitomi Soft X-ray Spectrometer
\thanks{The corresponding authors are
Masahiro \textsc{Tsujimoto},
Koji \textsc{Mori},
Shiu-Hang \textsc{Lee},
Hiroya \textsc{Yamaguchi},
Nozomu \textsc{Tominaga},
Takashi J. \textsc{Moriya},
Toshiki \textsc{Sato},
Cor de \textsc{Vries},
and
Ryo \textsc{Iizuka}
}
}
\author{Hitomi Collaboration,
Felix \textsc{Aharonian}\altaffilmark{1},
Hiroki \textsc{Akamatsu}\altaffilmark{2},
Fumie \textsc{Akimoto}\altaffilmark{3},
Steven W. \textsc{Allen}\altaffilmark{4,5,6},
Lorella \textsc{Angelini}\altaffilmark{7},
Marc \textsc{Audard}\altaffilmark{8},
Hisamitsu \textsc{Awaki}\altaffilmark{9},
Magnus \textsc{Axelsson}\altaffilmark{10},
Aya \textsc{Bamba}\altaffilmark{11,12},
Marshall W. \textsc{Bautz}\altaffilmark{13}
% ...
}
\altaffiltext{1}{Dublin Institute for Advanced Studies, 31 Fitzwilliam Place, Dublin 2, Ireland}
\altaffiltext{2}{SRON Netherlands Institute for Space Research, Sorbonnelaan 2, 3584 CA Utrecht, The Netherlands}
\altaffiltext{3}{Department of Physics, Nagoya University, Furo-cho, Chikusa-ku, Nagoya, Aichi 464-8601}
\altaffiltext{4}{Kavli Institute for Particle Astrophysics and Cosmology, Stanford University, 452 Lomita Mall, Stanford, CA 94305, USA}
\altaffiltext{5}{Department of Physics, Stanford University, 382 Via Pueblo Mall, Stanford, CA 94305, USA}
\altaffiltext{6}{SLAC National Accelerator Laboratory, 2575 Sand Hill Road, Menlo Park, CA 94025, USA}
\altaffiltext{7}{NASA, Goddard Space Flight Center, 8800 Greenbelt Road, Greenbelt, MD 20771, USA}
\altaffiltext{8}{Department of Astronomy, University of Geneva, ch. d'\'Ecogia 16, CH-1290 Versoix, Switzerland}
\altaffiltext{9}{Department of Physics, Ehime University, Bunkyo-cho, Matsuyama, Ehime 790-8577}
\altaffiltext{10}{Department of Physics and Oskar Klein Center, Stockholm University, 106 91 Stockholm, Sweden}
\altaffiltext{11}{Department of Physics, The University of Tokyo, 7-3-1 Hongo, Bunkyo-ku, Tokyo 113-0033}
\altaffiltext{12}{Research Center for the Early Universe, School of Science, The University of Tokyo, 7-3-1 Hongo, Bunkyo-ku, Tokyo 113-0033}
\altaffiltext{13}{Kavli Institute for Astrophysics and Space Research, Massachusetts Institute of Technology, 77 Massachusetts Avenue, Cambridge, MA 02139, USA}
% ...


\email{tsujimot@astro.isas.jaxa.jp, mori@astro.miyazaki-u.ac.jp}


\maketitle

\begin{abstract}
 The Crab nebula originated from a core-collapse supernova (SN) explosion observed in
 1054 A.\,D. When viewed as a supernova remnant (SNR), it has an anomalously low
 observed ejecta mass and kinetic energy for an Fe-core collapse SN. Intensive searches
 were made for a massive shell that solves this discrepancy, but none has been
 detected. An alternative idea is that the SN\,1054 is an electron-capture (EC)
 explosion with a lower explosion energy by an order of magnitude than Fe-core collapse
 SNe. In the X-rays, imaging searches were performed for the plasma emission from the
 shell in the Crab outskirts to set a stringent upper limit to the X-ray emitting
 mass. However, the extreme brightness of the source hampers access to its vicinity. We
 thus employed spectroscopic technique using the X-ray micro-calorimeter onboard the
 Hitomi satellite. By exploiting its superb energy resolution, we set an upper limit for
 emission or absorption features from yet undetected thermal plasma in the 2--12 keV
 range. We also re-evaluated the existing Chandra and XMM-Newton data. By assembling
 these results, a new upper limit was obtained for the X-ray plasma mass of $\lesssim$
 1~$M_{\odot}$ for a wide range of assumed shell radius, size, and plasma temperature
 both in and out of the collisional equilibrium. To compare with the observation, we
 further performed hydrodynamic simulations of the Crab SNR for two SN models (Fe-core
 versus EC) under two SN environments (uniform ISM versus progenitor wind). We found
 that the observed mass limit can be compatible with both SN models if the SN
 environment has a low density of $\lesssim$ 0.03~cm$^{-3}$ (Fe core) or $\lesssim$
 0.1~cm$^{-3}$ (EC) for the uniform density, or a progenitor wind density somewhat less
 than that provided by a mass loss rate of $10^{-5} M_{\odot}$~yr$^{-1}$ at
 20~km~s$^{-1}$ for the wind environment.
\end{abstract}

\section{Introduction}\label{s1}
Out of some 400\footnote{See http://www.physics.umanitoba.ca/snr/SNRcat/ for the
high-energy catalogues of SNRs and the latest statistics.} Galactic supernova remnants
(SNRs) detected in the X-rays and $\gamma$-rays \citep{ferrand12}, about 10\% of them
lack shells, which is one of the defining characteristics of SNRs. They are often
identified instead as pulsar wind nebulae (PWNe), systems that are powered by the
rotational energy loss of a rapidly rotating neutron star generated as a consequence of
a core-collapse supernova (SN) explosion.

The lack of a shell in these sources deserves wide attention, since it is a key to
unveiling the causes behind the variety of observed phenomena in SNRs. In this pursuit,
it is especially important to interpret in the context of the evolution from SNe to
SNRs, not just a taxonomy of SNRs. Observed results of SNRs do exhibit imprints of their
progenitors, explosion mechanisms, and surrounding environment
\citep{hughes95,yamaguchi14b}. Recent rapid progress in simulation studies of the stellar
evolution of progenitors, SN explosions, and hydrodynamic development of SNRs makes it
possible to gain insights about SNe from SNR observations.

\medskip

The Crab nebula is one such source. It is an observational standard for X-ray and
$\gamma$-ray flux and time
\citep{kirsch05,kaastra09,weisskopf10,madsen15,jahoda06,terada08}. As a PWN, the Crab
exhibits typical X-ray and $\gamma$-ray luminosities for its spin-down luminosity
\citep{possenti02,kargaltsev12,mattana09} and a typical morphology
\citep{ng08,bamba10}. It also played many iconic roles in the history of astronomy, such
as giving observational proof \citep{staelin68,lovelace68} for the birth of a neutron
star in SN explosions \citep{baade34} and linking modern and ancient astronomy by its
association with a historical SN in 1054 documented primarily in Oriental records
\citep{stephenson02,lundmark21,rudie08}.

This astronomical icon, however, is known to be anomalous when viewed as an SNR. Besides
having no detected shell, it has an uncomfortably small observed ejecta mass of 4.6
$\pm$ 1.8 $M_{\odot}$ \citep{fesen97}, kinetic energy of $\lesssim 1 \times 10^{50}$ erg
\citep{davidson85}, and maximum velocity of only 2,500~km~s$^{-1}$ \citep{sollerman00},
all of which are far below the values expected for a typical core-collapse SN.

One idea to reconcile this discrepancy is that there is a fast and thick shell yet to be
detected, which carries a significant fraction of the mass and kinetic energy
\citep{chevalier77}. If the free expansion velocity is 10$^{4}$~km~s$^{-1}$, the shell
radius has grown to 10~pc over 10$^{3}$~yr. Intensive attempts were made to detect such
a shell in the radio \citep{frail95}, H$\alpha$ \citep{tziamtzis09}, and X-rays
\citep{mauche85,predehl95,seward06}, but without success.

Another idea is that the SN explosion was indeed anomalous to begin with.
\citet{nomoto82} proposed that SN\,1054 was an electron-capture (EC) SN, which is caused
by the endothermic reaction of electrons captured in an O-Ne-Mg core, in contrast to the
photo-dissociation in an Fe core for the normal core-collapse SN. EC SNe are considered
to be caused by an intermediate (8--10 $M_{\odot}$) mass progenitor in the asymptotic
giant branch (AGB) phase. Simulations based on the first principle calculation
\citep{kitaura06,janka08} show that an explosion takes place with a small energy of
$\sim$10$^{50}$~erg, presumably in a dense circumstellar environment as a result of the
mass loss by a slow but dense stellar wind. This idea matches well with the
aforementioned observations of the Crab, plus the richness of the He abundance
\citep{macalpine08}, an extreme brightness in the historical records
\citep{sollerman01,tominaga13,moriya14}, and the observed nebular size
\citep{yang15}. If this is the case, we should rather search for the shell much closer
to the Crab.

\medskip

The X-ray band is most suited to search for the thermal emission from a 10$^{6}$--10$^{8}$~K
plasma expected from the shocked material forming a shell. In the past, telescopes with
a high spatial resolution were used to set an upper limit on the thermal X-ray emission from the
Crab \citep{mauche85,predehl95,seward06}. A high contrast imaging is required to
minimize the contamination by scattered X-rays by the telescope itself and the
interstellar dust around the Crab. Still, the vicinity of the Crab is inaccessible with
the imaging technique for the overwhelmingly bright and non-uniform flux of the PWN.

Here, we present the result of a spectroscopic search for the thermal plasma using the
soft X-ray spectrometer (SXS) onboard the Hitomi satellite
\citep{takahashi16}. The SXS is a non-dispersive high-resolution spectrometer, offering
a high contrast spectroscopy to discriminate the thermal emission or absorption lines
from the bright featureless spectrum of the PWN. This technique allows access to the
Crab's vicinity and is complementary to the existing imaging results.

The goals of this paper are (1) to derive a new upper limit with the spectroscopic
technique for the X-ray emitting plasma, (2) to assemble the upper limits by various
techniques evaluated under the same assumptions, and (3) to compare with the latest
hydro-dynamic (HD) calculations to examine if any SN explosion and environment models
are consistent with the X-ray plasma limits. We start with the observations and the data
reduction of the SXS in \S~\ref{s2}, and present the spectroscopic search results of
both the absorption and emission features by the thermal plasma in \S~\ref{s3}. In
\S~\ref{s4}, we derive the upper limits on the physical parameters of the SN and the SNR
using our results presented here and existing result in the literature, and compare with
our HD simulations to gain insight into the origin of SN\,1054.

\section{Observations and Data Reduction}\label{s2}
\subsection{Observations}\label{s2-1}
The SXS is a high-resolution X-ray spectrometer based on X-ray micro-calorimetry
\citep{kelley16}. The HgTe absorbers placed in a 6$\times$6 array absorb individual
X-ray photons collected by the X-ray telescope, and the temperature increase of the Si
thermometer is read out as a change in its resistance. Because of the very low heat
capacity of the sensor controlled at a low temperature of 50~mK, a high spectral
resolution is achieved over a wide energy range. The SXS became the first X-ray
micro-calorimeter to have made observations of astronomical sources in the orbit and
proved its excellent performance despite its short lifetime.

The Crab was observed on 2016 March 25 from 12:35 to 18:01 UT with the SXS. This turned
out to be the last data set collected before the tragic loss of the spacecraft on the
next day. The observation was performed as a part of the calibration program, and we
utilize the data to present scientific results in this paper.

\begin{figure}
 \begin{center}
  \includegraphics[width=0.5\textwidth]{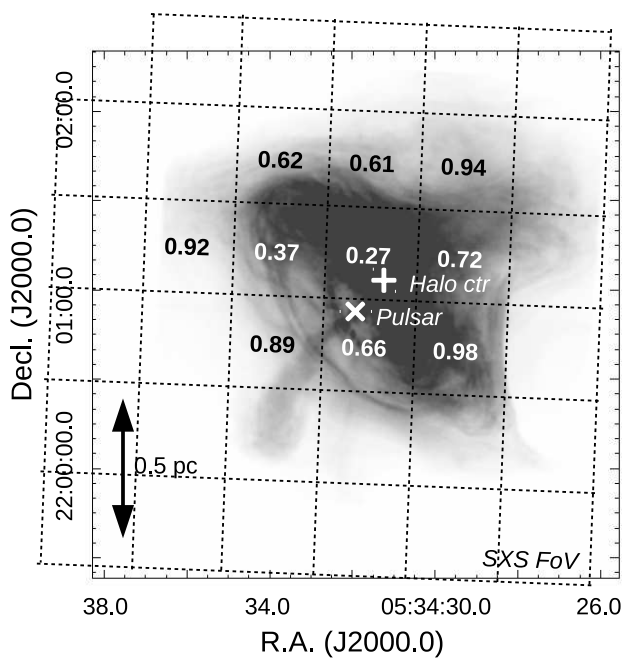}
 \end{center}
 \vspace*{8mm}
 \caption{Field of view of the SXS superposed on the Chandra ACIS image after correcting for
 the readout streaks \citep{mori04}. The 6$\times$6 pixels are shown with the top left
 corner uncovered for the calibration pixel. The numbers indicate the live time fraction
 only for pixels less than 0.980. The astrometry of the SXS events can be displaced by
 20\farcs6 at 1 $\sigma$ when the star tracker is unavailable. The position of the
 pulsar \citep{lobanov11} and the halo center \citep{seward06} are respectively shown
 with the cross and the plus signs.}
 \label{f10}
\end{figure}

Figure~\ref{f10} shows the 3\farcm0 $\times$ 3\farcm0 field of view on top of a Chandra
image. The scale corresponds to 1.9~pc at a distance of 2.2~kpc
\citep{manchester05}. This covers a significant fraction of the observed elliptical
nebula with a diameter of 2.9$\times$4.4~pc \citep{hester08}. The SXS was still in the
commissioning phase \citep{tsujimoto16}, and some instrumental setups were
non-nominal. Among them, the gate valve status was most relevant for the result
presented here. The valve was closed to keep the Dewar in a vacuum on the ground, which
was planned to be opened when we confirmed the initial outgassing had ceased in the
spacecraft. This observation was made before this operation. As a result, the
attenuation by a $\sim$260~$\mu$m Be window of the gate valve \citep{eckart16} limited
the SXS bandpass to above $\sim$2~keV, which would otherwise extend down to
$\sim$0.1~keV. 

The instrument had reached the thermal equilibrium by the time of the observation
\citep{fujimoto16,noda16}. The detector gain was very stable except for the passage of
the South Atlantic anomaly. The previous recycle operation of the adiabatic
demagnetization refrigerators was started well before the observation at 10:20 on March
24, and the entire observation was within its 48-hour hold time \citep{shirron16}. The
energy resolution was 4.9~eV measured with the $^{55}$Fe calibration source at 5.9~keV
for the full width at the half maximum \citep{porter16,kilbourne16b,leutenegger16}. This
superb resolution is not compromised by the extended nature of the Crab nebula for being
a non-dispersive spectrometer.

The actual incoming flux measured with the SXS was equivalent to $\sim$0.3~Crab in the
2--12 keV band due to the extra attenuation by the gate valve. The net exposure time was
9.7~ks.

\subsection{Data Reduction}\label{s2-2}
We started with the cleaned event list produced by the pipeline process version
03.01.005.005 \citep{angelini16}. Throughout this paper, we used the HEASoft and CALDB
release on 2016 December 22 for the Hitomi collaboration. Further screening against
spurious events was applied based on the energy versus pulse rise time. The screening
based on the time clustering of multiple events was not applied; it is intended to
remove events hitting the out-of-pixel area, but a significant number of false positive detection is
expected for high count rate observations like this.

Due to the high count rate, some pixels at the array center suffer dead time
(figure~\ref{f10}; \cite{ishisaki16}). Still, the observing efficiency of $\sim$72\% for
the entire array is much higher than conventional CCD X-ray spectrometers. For example,
Suzaku XIS \citep{koyama07} requires a 1/4 window $+$ 0.1~s burst clocking mode to avoid
pile-up for a 0.3~Crab source, and the efficiency is only $\sim$5\%. Details of the dead
time and pile-up corrections are described in a separate paper. We only mention here
that these effects are much less serious for the SXS than CCDs primarily due to a much
faster sampling rate of 12.5~kHz and a continuous readout.

The source spectrum was constructed in the 2--12~keV range at a resolution of
0.5~eV~bin$^{-1}$. Events not contaminated by other events close in time (graded as Hp
or Mp; \cite{kelley16}) were used for a better energy resolution. All pixels were
combined. The redistribution matrix function was generated by including the energy loss
processes by escaping electrons and fluorescent X-rays. The half power diameter of the
telescope is 1\farcm2 \citep{okajima16}. The SXS has only a limited imaging capability,
and we do not attempt to perform a spatially-resolved spectroscopic study in this
paper. The SXS does have a timing resolution to resolve the 34~ms pulse phase, but we do
not attempt a phase-resolved study either as only a small gain in the contrast of
thermal emission against the pulse emission is expected; the unpulsed emission of a
$\sim$90\% level of averaged count rate can be extracted at a compensation of $\sim$2/3
of the exposure time.

The total number of events in the 2--12 keV range is 7.6$\times$10$^{5}$. The background
spectrum, which is dominated by the non-X-ray background, was accumulated using the data
when the telescope was pointed toward the Earth. The non-X-ray background is known to
depend on the strength of the geomagnetic field strength at the position of the
spacecraft within a factor of a few. The history of the geomagnetic cut-off rigidity
during the Crab observation was taken into consideration to derive the background rate
as 8.6$\times$10$^{-3}$ s$^{-1}$ in the 2--12~keV band. This is negligible with
$\sim$10$^{-4}$ of the source rate.

\section{Analysis}\label{s3}
To search for signatures of thermal plasma, we took two approaches. One is to add a
thermal plasma emission model, or to multiply a thermal plasma absorption model, upon
the best-fit continuum model with an assumed plasma temperature, which we call plasma
search (\S~\ref{s3-1}). Here, we assume that the feature is dominant either as emission
or absorption. The other is a blind search of emission or absorption lines, in which we
test the significance of an addition or a subtraction of a line model upon the best-fit
continuum model (\S~\ref{s3-2}). For the spectral fitting, we used the \texttt{Xspec}
package version 12.9.0u \citep{arnaud96}. The statistical uncertainties are evaluated at
1$\sigma$ unless otherwise noted.

\subsection{Plasma search}\label{s3-1}
\subsubsection{Fiducial model}\label{s3-1-1}
We first constructed the spectral model for the entire energy band. The spectrum was
fitted reasonably well with a single power-law model with an interstellar extinction,
which we call the fiducial model. Hereafter, all the fitting was performed for unbinned
spectra based on the C statistics \citep{cash79}. For the extinction model by cold
matter, we used the \texttt{tbabs} model version 2.3.2\footnote{See
http://pulsar.sternwarte.uni-erlangen.de/wilms/research/tbabs/ for details.}
\citep{wilms00}. We considered the extinction by interstellar gas, molecules, and dust
grains with the parameters fixed at the default values of the model except for the total
column density. The SXS is capable of resolving the fine structure of absorption edges,
which is not included in the model except for O K, Ne K, and Fe L edges. This, however,
does not affect the global fitting, as the depths of other edges are shallow for the Crab
spectrum.

\begin{figure}
 \begin{center}
  \includegraphics[width=0.45\textwidth]{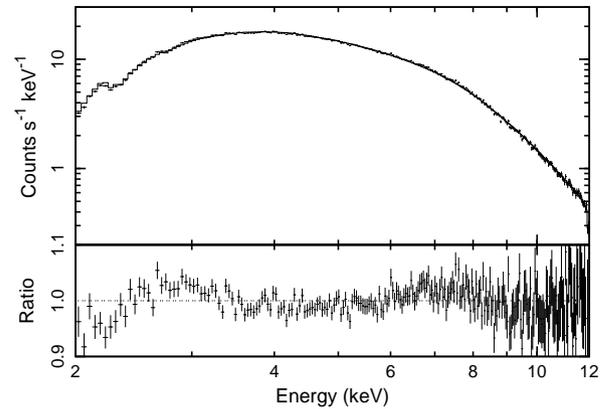}
 \end{center}
 \vspace*{8mm}
 \caption{Best-fit fiducial model to the background-subtracted spectra binned only for
 display purpose. The top panel shows the data with crosses and the best-fit model with
 solid lines. The bottom panel shows the ratio to the fit.}
 \label{f01}
\end{figure}

We calculated the effective area assuming a point-like source at the center of the SXS
field. The nebula size is no larger than the point spread function. Figure~\ref{f01}
shows the best-fit model, while table~\ref{t01} summarizes the best-fit parameters for
the extinction column by cold matter ($N_{\mathrm{H}}^{\mathrm{(cold)}}$), the power-law
photon index ($\Gamma$), and the X-ray flux ($F_{\mathrm{X}}$). The ratio of the data to
the model show some broad features, which are attributable to the inaccuracies of the
calibration including the mirror Au M and L edge features, the gate valve transmission,
the line spread function, ray-tracing modeling accuracies, etc (Okajima et al. in
prep.). In this paper, therefore, we constrain ourselves to search for lines that are
sufficiently narrow to decouple with these broad systematic uncertainties. This is
possible only with high-resolution spectrometers.

\begin{table}[ht]
 \caption{Best-fit parameters of the global fitting.}
 \label{t01}
 \begin{center}
  \begin{tabular}{lc}
   \hline
   \hline
   Parameter\footnotemark[$*$]       & Best-fit \\
   \hline
   $N_{\mathrm{H}}^{\mathrm{(cold)}}$ 10$^{21}$~cm$^{-2}$  & 4.6 (4.1--5.0) \\
   $\Gamma$                              & 2.17 (2.16--2.17) \\
   $F_{\mathrm{X}}$ erg~s$^{-1}$~cm$^{-2}$\footnotemark[$\dagger$] & 1.722 (1.719--1.728) $\times$10$^{-8}$ \\
   Red-$\chi^{2}$/d.o.f.             & 1.34/19996 \\
   \hline
   \multicolumn{2}{l}{\parbox{60mm}{
   \footnotesize
   \par \noindent
   \footnotemark[$*$] The errors indicate a 1$\sigma$ statistical uncertainty. \\
   \footnotemark[$\dagger$] The absorption-corrected flux at 2--8~keV.\\
   }}
  \end{tabular}
 \end{center}
\end{table}

\subsubsection{Plasma emission}\label{s3-1-2}
For the thermal plasma emission, we assumed the optically-thin collisional ionization
equilibrium (CIE) plasma model and two non-CIE deviations from it. All the calculations
were based on the atomic database \texttt{ATOMDB} \citep{foster12} version 3.0.7. We
assumed the solar abundance \citep{wilms00}. This gives a conservative upper limit for
plasma with a super-solar metalicity when they are searched using metallic lines.

\begin{figure}
 \begin{center}
  \includegraphics[width=0.5\textwidth]{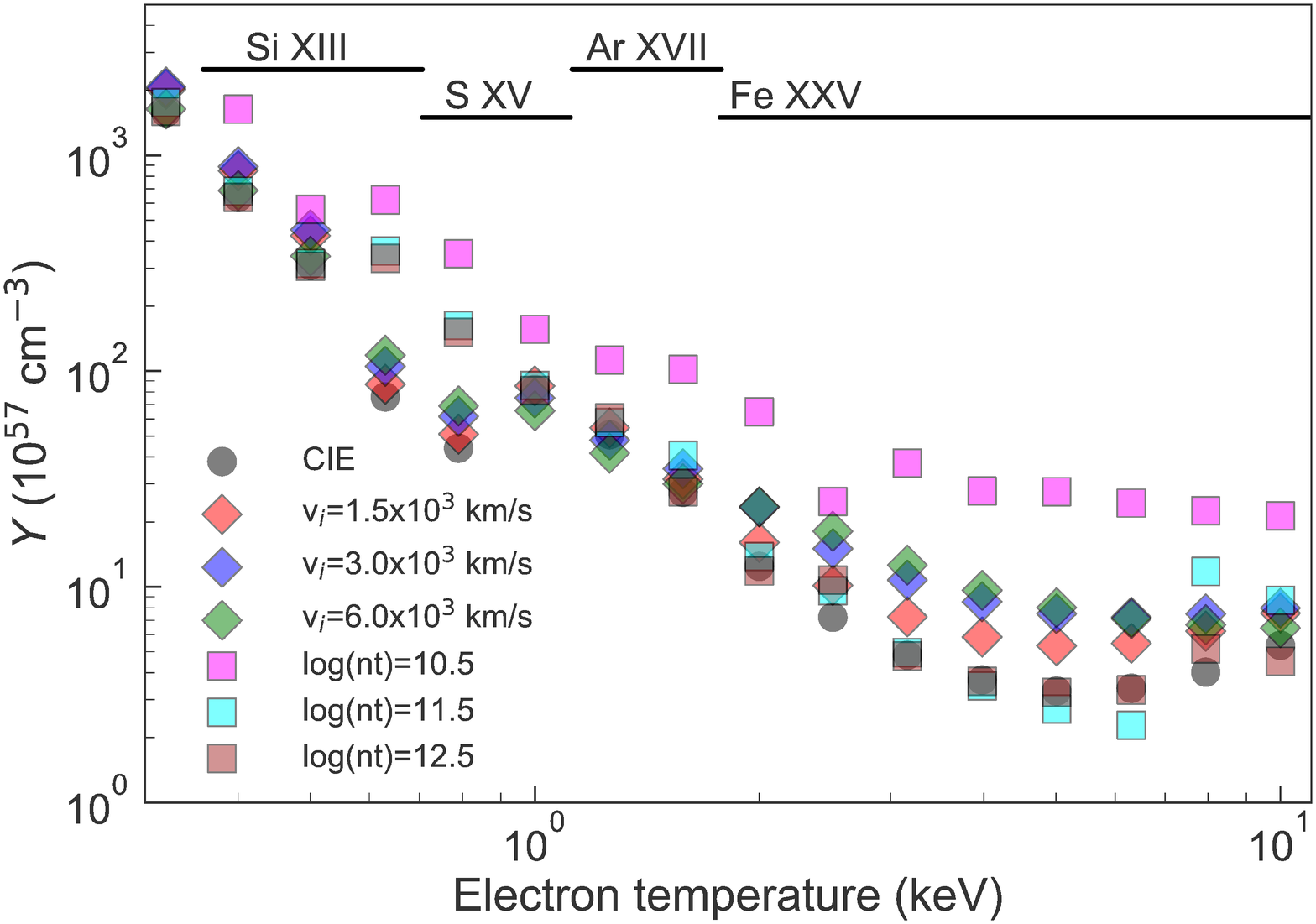}
 \end{center}
 \caption{3$\sigma$ statistical upper limits of the volume emission measure ($Y$) for the
 assumed electron temperature for selected parameters (table~\ref{t02}): (a) CIE, (b)
 broadened lines by $v_{i}=$(1.5, 3.0, and 6.0) $\times$10$^{3}$~km~s$^{-1}$, and (c)
 non-equilibrium cases with $\log{(nt~\mathrm{cm^{-3}~s})}=$10.5, 11.5, and 12.5. The
 name of ions giving the strongest emission line for (a) at each temperature is shown at
 the top.}
\label{f05}
\end{figure}

First, we used the \texttt{apec} model \citep{smith01} for the CIE plasma, in which the
electron, ion, and ionization temperatures are the same. Neither the bulk motion nor the
turbulence broadening was considered, but the thermal broadening was taken into account
for the lines. For each varying electron temperature (table~\ref{t02}), we selected the
strongest emission line in the 10 non-overlapping 1 keV ranges in the 2--12~keV
band. For each selected line, we first fitted the $\pm$50~eV range around the line with
a power-law model, then added the plasma emission model to set the upper limit of the
volume emission measure ($Y$) of the plasma. Both power-law and plasma emission models
were attenuated by an interstellar extinction model of a column density fixed at the
fiducial value (table~\ref{t01}). We expect some systematic uncertainty in the
$N_{\mathrm{H}}^{\mathrm{(cold)}}$ value due to incomplete calibration at low
energies. The best-fit value in the fiducial model (table~\ref{f01}) tends to be higher
than those in the literature \citep{kaastra09,weisskopf10} by 10--30\%. A 10\% decrease
in the value leads to $<$10\% decrease of $Y$ for the temperature $>$1~keV. The
normalization of the plasma model was allowed to vary both in the positive and negative
directions so as not to distort the significance distribution. The result for selected
cases is shown in figure~\ref{f05}.

Deviation from the thermal equilibrium is seen in SNR plasmas
\citep{borkowski01,vink12}, especially for young SNRs expanding in a low density
environment. We considered two types of deviations. One is the non-equilibrium
ionization using the \texttt{nei} model \citep{smith10}. This code calculates the
collisional ionization as a function of the ionization age ($n_{\mathrm{e}}t$), and
accounts for the difference between the ionization and electron temperatures. The
electron temperature is assumed constant, which is reasonable considering that some SNRs
show evidence for the collision-less instantaneous electron heating at the shock
\citep{yamaguchi14a}. We took the same procedure with the CIE plasma for the
$n_{\mathrm{e}}t$ values listed in table~\ref{t02}, and derived the upper limit of $Y$.

Another non-CIE deviation is that the electron and ion temperatures are different. More
massive ions are expected to have a higher temperature than less massive ions and
electrons, hence are more thermally broadened before reaching equilibrium. We derived
the upper limit of $Y$ for several values of the ion's thermal velocity $v_{i}$
(table~\ref{t02}). In this modeling, the continuum fit was performed over an energy
range of the smaller of the two: $\pm$(3$\times E v_{i}/c$ or 50)~eV centered at the
line energy $E$, so as to decouple the continuum and line fitting when $v_{i}$ is large.

\begin{table*}[ht]
 \caption{Investigated parameter space.}
 \label{t02}
 \begin{center}
  \begin{tabular}{lllrl}
   \hline
   \hline
   Par & Unit & Description & Total\footnotemark[$\S$] & Cases\footnotemark[$\S$] \\
   \hline
   $T_{\mathrm{e}}$ & keV  & Electron temperature & 21 & 0.1--10 (0.1 dex step) \\
   $\log{(n_{\mathrm{e}}t)}$\footnotemark[$*$] & s cm$^{-3}$ & Ionization age & 8 & 10.0--13.5 (0.5 step) \\
   $v_{\mathrm{i}}/c$\footnotemark[$*\dagger$] & & Thermal broadening of lines & 5 & 0.001, 0.002, 0.005, 0.01, 0.02 \\
   $\Delta R/R$\footnotemark[$\ddagger$] & & Shell fraction & 6 & 0.005, 0.01, 0.05,
		   0.083 ($=$1/12), 0.10, 0.15 \\
   \hline
   \multicolumn{5}{l}{\parbox{160mm}{
   \footnotesize
   \par \noindent
   \footnotemark[$*$] The parameter is searched only for the plasma emission (\S~\ref{s3-1-2}).\\
   \footnotemark[$\dagger$] The ion spices $i$ has a velocity $v_{i}$, thus has a
   temperature of $T_{i} = m_{i}v_{i}^{2}$/$k_{\mathrm{B}}$, in which $m_{i}$ is the mass of the
   ion. In the case of Si and Fe, the cases correspond to $T_{\mathrm{Si}}<$ 12~MeV and $T_{\mathrm{Fe}}<$ 21 MeV.\\
   \footnotemark[$\ddagger$] The value 1/12 is for the self-similar solution
   \citep{sedov59}, and 0.15 follows preceding work \citep{seward06,frail95}.\\
   \footnotemark[$\S$] The adopted parameters (cases) and the total number of cases
   (total) are shown.\\
   }}
  \end{tabular}
 \end{center}
\end{table*}

\subsubsection{Plasma absorption}\label{s3-1-3}
\begin{figure}
 \begin{center}
  \includegraphics[width=0.5\textwidth]{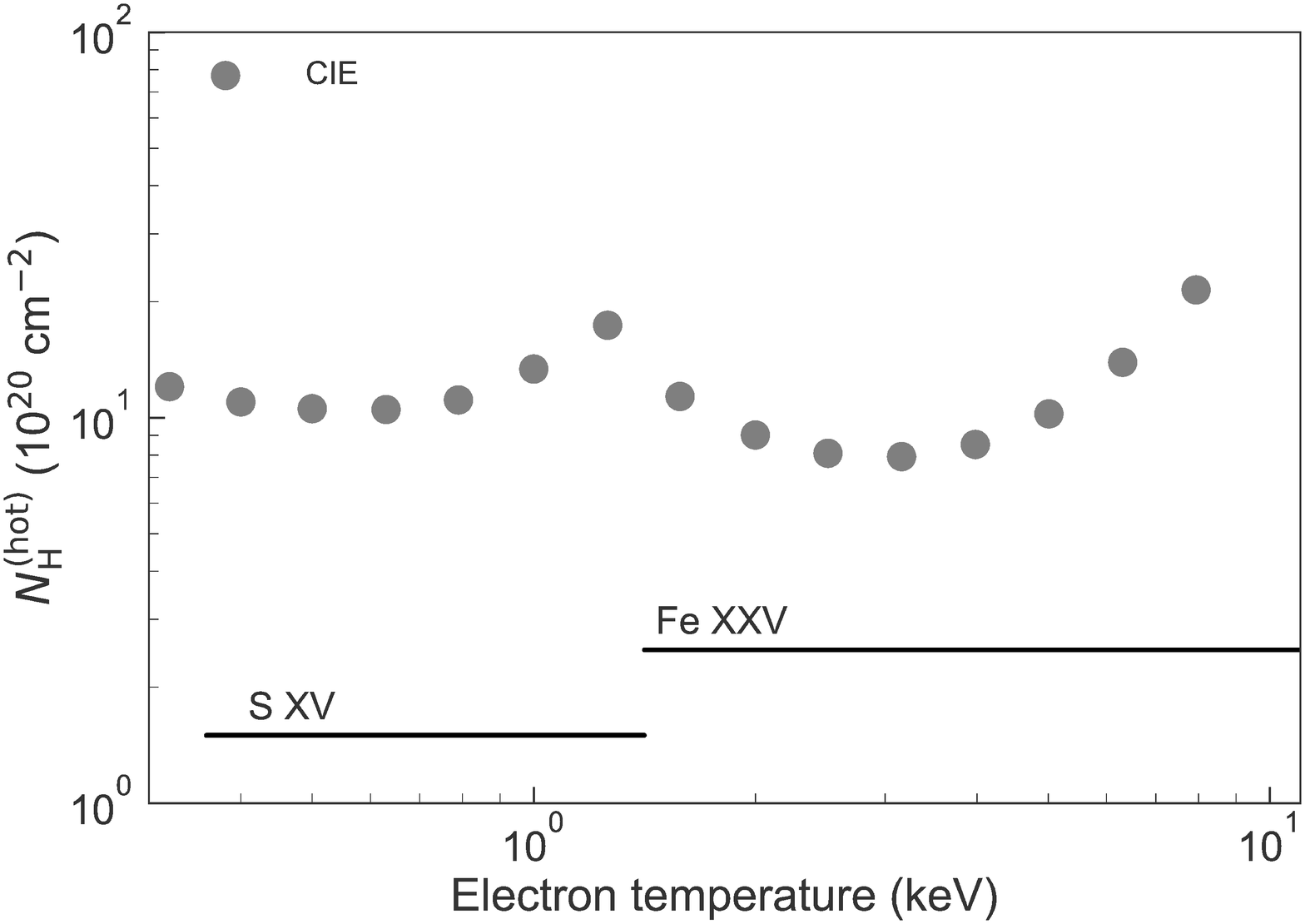}
 \end{center}
 \caption{3$\sigma$ statistical upper limits of the hydrogen-equivalent extinction column
 ($N_{\mathrm{H}}^{\mathrm{(hot)}}$) by the CIE plasma for the assumed electron
 temperature. The name of ions giving the strongest absorption line at each temperature
 is shown at the bottom}
 \label{f06}
\end{figure}

A similar procedure was taken for deriving the upper limit for the absorption column by
a thermal plasma. We used the \texttt{hotabs} model \citep{kallman01} and only
considered the CIE plasma. At each assumed electron temperature (table~\ref{t02}), we
selected the strongest absorption line in the 10 non-overlapping 1 keV ranges in the
2--12 keV band. For each selected line, we first fitted the $\pm$50 eV range around the
line with a power-law model, then multiplied the plasma absorption model to set the
upper limit of the hydrogen-equivalent absorption column
($N_{\mathrm{H}}^{\mathrm{(hot)}}$) by the plasma. The result is shown in
figure~\ref{f06}.

\subsubsection{Example in the Fe K band}\label{s3-1-4}
For the emission, the resultant upper limit of $Y$ is less constrained for plasma with
lower temperatures. At low temperatures, strong lines are at energies below 2~keV, in
which the SXS has no sensitivity as the gate valve was not opened. For increasing
temperatures above $\sim$0.5~keV, S-He$\alpha$, Ar-He$\alpha$, or Fe-He$\alpha$ are used
to set the limit. The most stringent limit is obtained at the maximum formation
temperature ($\sim$5~keV) of the Fe-He$\alpha$ line. For the NEI plasma with a low
ionization age (10$^{10.5}$~s~cm$^{-3}$), He-like Fe ions have not been formed yet, thus
the limit is not stringent. Conversely, at an intermediate ionization age
(10$^{11.5}$~s~cm$^{-3}$), Fe is not fully ionized yet, thus Fe-He$\alpha$ can give a
strong upper limit even for electron temperatures of $\sim$10~keV. At
10$^{12.5}$~s~cm$^{-3}$, the result is the same with the CIE plasma as expected.

\begin{figure}
 \begin{center}
  \includegraphics[width=0.5\textwidth]{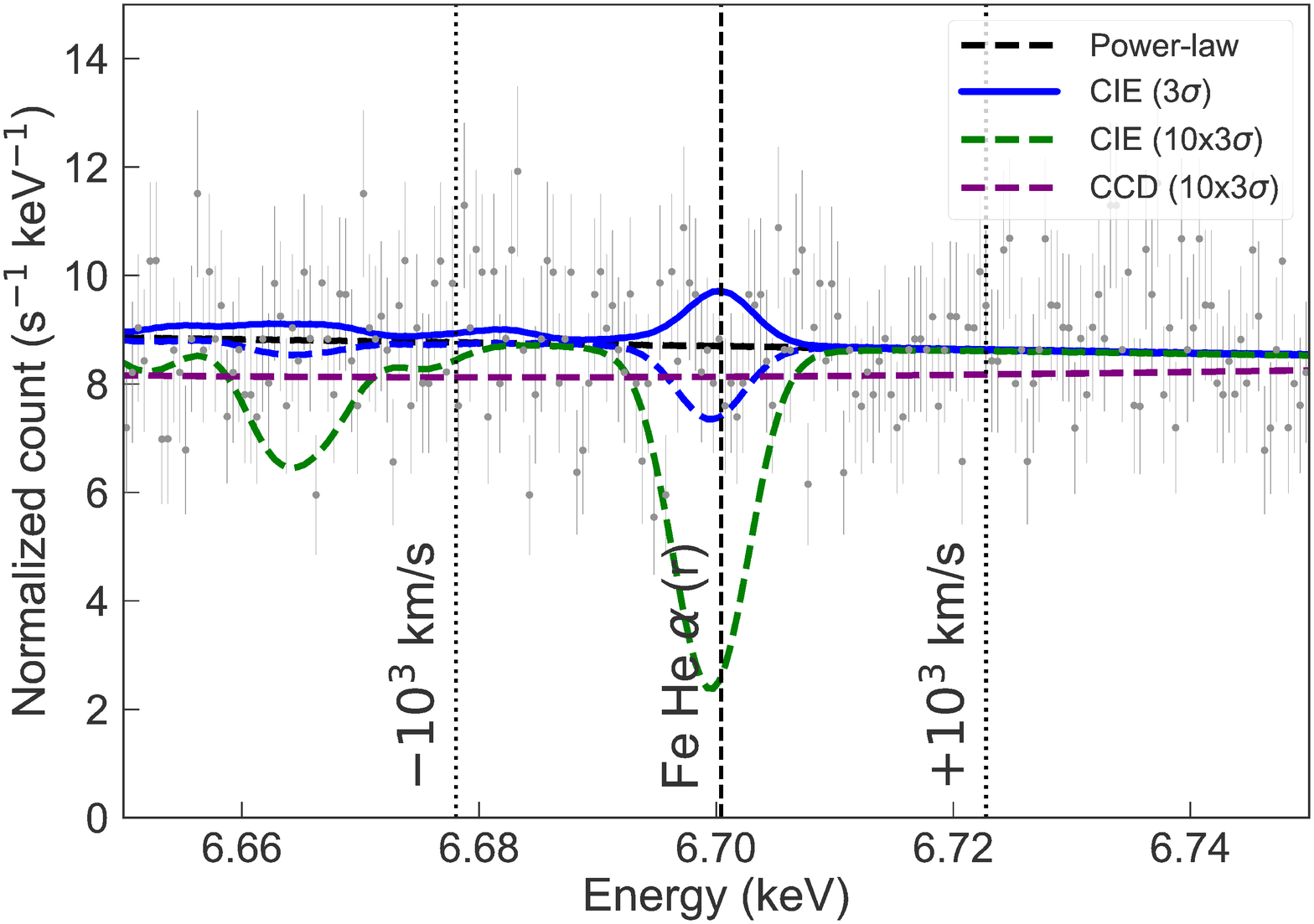}
 \end{center}
 \caption{Close-up view around the Fe-He$\alpha$ resonance line. Over the unbinned
 spectrum (gray plus signs), several models are shown: the best-fit continuum model (black
 dashed), and the emission (solid) and absorption (dashed) by a 3.16~keV CIE plasma with
 3$\sigma$ upper limits (blue) corresponding to $Y=2.1 \times 10^{57}$~cm$^{-3}$ for
 emission and $N_{\mathrm{H}}^{\mathrm{(hot)}}=7.9 \times 10^{20}$~cm$^{-2}$ for
 absorption. Ten times the absorption value is also shown with green (SXS) and purple
 (convolved with a Suzaku XIS response).}
 \label{f04}
\end{figure}

Figure~\ref{f04} shows a close-up view of the fitting around the Fe-He$\alpha$ line for
the case of the 3.16~keV electron temperature. Overlaid on the data, models are shown in
addition to the best-fit power-law continuum model. Also shown is the expected result by
a CCD spectrometer, with which the levels detectable easily with the SXS would be
indistinguishable from the continuum emission. This demonstrates the power of an X-ray
micro-calorimeter for weak features from extended sources. The expected energy shifts
for a bulk velocity of $\pm$10$^{3}$~km~s$^{-1}$, or $\pm$22.4~eV, are shown. The data
quality is quite similar in this range, thus the result is not significantly affected by
a possible gain shift ($\lesssim$1~eV; \cite{hitomi16}) or a single bulk velocity shift.

\subsection{Blind search}\label{s3-2}
We searched for emission or absorption line features at an arbitrary line energy in the
2--12~keV range. We made trials at 20,000 energies separated by 0.5~eV. The trials were
repeated for a fixed line width corresponding to a velocity of $v=$ 0, 20, 40, 80, 160,
320, 640, and 1280~km~s$^{-1}$. For each set of line energy and width, we fitted the
spectrum with a power-law model locally in an energy range 3--20 $\sigma_{E}(E)$ on both
sides of the trial energy $E$. Here, the unit of the fitting range $\sigma_{E}(E)$ is
determined as
\begin{equation}
 \sigma_E (E)= \sqrt{ (E (v/c))^{2} + (\Delta E_{\mathrm{det}}(E))^{2} },
\end{equation}
in which $\Delta E_{\mathrm{det}}(E)$ is the 1$\sigma$ width of the Gaussian core of the
detector response \citep{leutenegger16}. With this variable fitting range, we can test a
wide range of line energy and width. After fixing the best-fit power-law model, we added
a Gaussian model allowing both positive and negative amplitudes respectively for
emission and absorption lines and refitted in the 0--20 $\sigma_{E}$ on both sides. The
detection significance was evaluated as
\begin{equation}
 \sigma = \frac{N_{\mathrm{line}}}{\sqrt{\Delta {N_{\mathrm{line}}^2} +
  (N_{\mathrm{line}} \Delta I_{\mathrm{cont}}/I_{\mathrm{cont}})^2}}\label{e01},
\end{equation}
in which $N_{\mathrm{line}}$ and $\Delta N_{\mathrm{line}}$ are the best-fit and
1$\sigma$ statistical uncertainty of the line normalization in the unit of
s$^{-1}$~cm$^{-2}$, whereas $I_{\mathrm{line}}$ and $\Delta I_{\mathrm{line}}$ are those
of the continuum intensity in the unit of s$^{-1}$~cm$^{-2}$~keV$^{-1}$ at the line
energy.

\begin{figure}
 \begin{center}
  \includegraphics[width=0.5\textwidth]{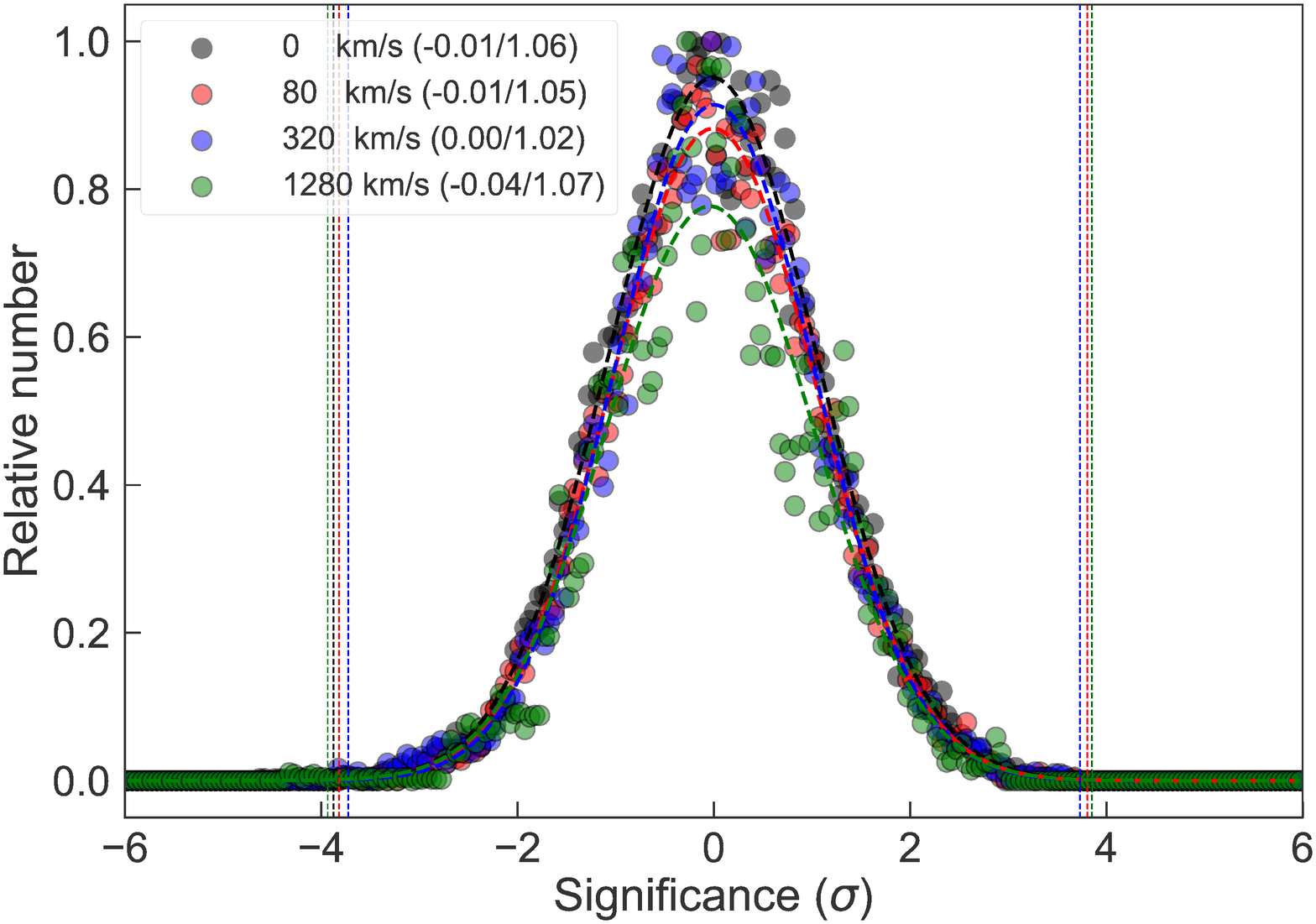}
 \end{center}
 \caption{Distribution of significance (eqn \ref{e01}) for different assumed velocities
 in different colors. The distribution is fitted by a single Gaussian model, and its
 best-fit parameters are shown in the legend as (center/width). The vertical dotted
 lines indicate the 5 $\sigma$ level of the best-fit Gaussian distribution.}
\label{f02}
\end{figure}

Figure~\ref{f02} shows the distribution of the significance. All are reasonably well
fitted by a single Gaussian distribution. We tested several different choices of fitting
ranges and confirmed that the overall result does not change. Above a 5$\sigma$ level
(0.01 false positives expected for 20,000 trials) of the best-fit Gaussian distribution,
no significant detection was found except for (1) several detections of absorption in the
2.0--2.2~keV energy range for a wide velocity range, and (2) a detection of absorption
at $\sim$9.48~keV for 160 and 320~km~s$^{-1}$. The former is likely due to the
inaccurate calibration of the Au M edges of the telescope. For the latter, no instrumental
features or strong atomic transitions are known around this energy. However, we do not
consider this to be robust as it escapes detection only by changing the fitting ranges.

\begin{figure}
 \begin{center}
  \includegraphics[width=0.5\textwidth]{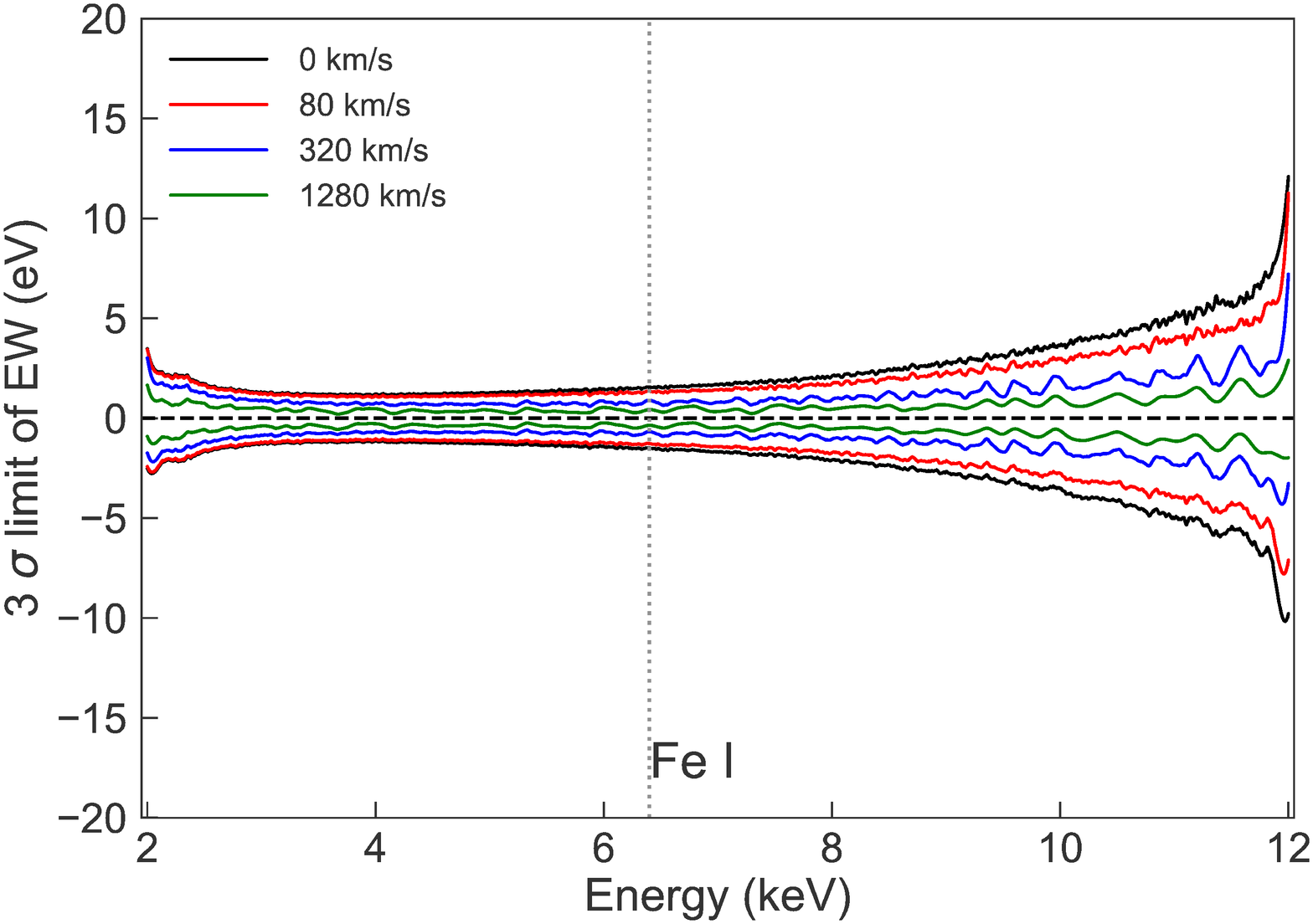}
 \end{center}
 \caption{3 $\sigma$ range of the equivalent width for different assumed
 velocities. The curves are obtained by convolving the fitting result at each energy bin
 with a low pass filter. A structure at 11.9~keV is due to the Au L$\alpha_{3}$
 absorption edge by the telescope.}
\label{f03}
\end{figure}

The equivalent width,
\begin{math}
\mathrm{EW} = N_{\mathrm{line}} / I_{\mathrm{cont}},
\end{math}
was derived for every set of the line energy and width along with their 3 $\sigma$
statistical uncertainty (figure~\ref{f03}). The 3$\sigma$ limit of EW at 6.4~keV is
$\lesssim$ 2~eV. We would expect the Fe fluorescence line with
\begin{math}
 \mathrm{EW} = \alpha (\Delta \Omega/4\pi) (N_{\mathrm{H}}^{\prime} / \mathrm{10^{22}~cm^{-2}})~\mathrm{eV},
\end{math}
in which $\alpha \sim 2.8$ for the Crab's power-law spectrum \citep{krolik87}. $\Delta
\Omega$ and $N_{\mathrm{H}}^{\prime}$ are, respectively, the subtended angle and the
H-equivalent column of the fluorescing matter around the incident emission. Assuming 
$\Delta \Omega = 4 \pi$ and $N_{\mathrm{H}}^{\prime} < 0.32 \times 10^{22}$~cm$^{-2}$,
which is the measured value in the line of sight inclusive of the ISM \citep{mori04},
the expected EW is consistent with the upper limit by the SXS.

\section{Discussion}\label{s4}
In \S~\ref{s4-1}, we convert the upper limit of $Y$ or $N_{\mathrm{H}}^{\mathrm{(hot)}}$
with the SXS into that of the plasma density ($n_{\mathrm{X}}$) by making several
assumptions. In \S~\ref{s4-2}, we re-evaluate the data by other methods in the literature
under the same assumptions to assemble the most stringent upper limit of
$n_{\mathrm{X}}$ for various ranges of the parameters. In \S~\ref{s4-3}, we perform a HD
calculation for some SN models and verify that the searched parameter ranges are
reasonable. In \S~\ref{s4-4}, we compare the HD result with observed limits.

\subsection{Constraints on the plasma density with SXS}\label{s4-1}
For converting the upper limits of $Y$ and $N_{\mathrm{H}}^{\mathrm{(hot)}}$ of the
thermal plasma into that of the X-ray emitting plasma density ($n_{\mathrm{X}}$), we
assume the plasma is uniform in a spherically symmetric shell in a range of $R$ to
$R+\Delta R$ from the center. We assumed several shell fraction ($\Delta R/R$) values
(table~\ref{t02}). For simplicity, the electron and ion densities are the same, and all
ions are hydrogen. This gives a conservative upper limit for the plasma mass.

\begin{figure}
 \begin{center}
  \includegraphics[width=0.5\textwidth]{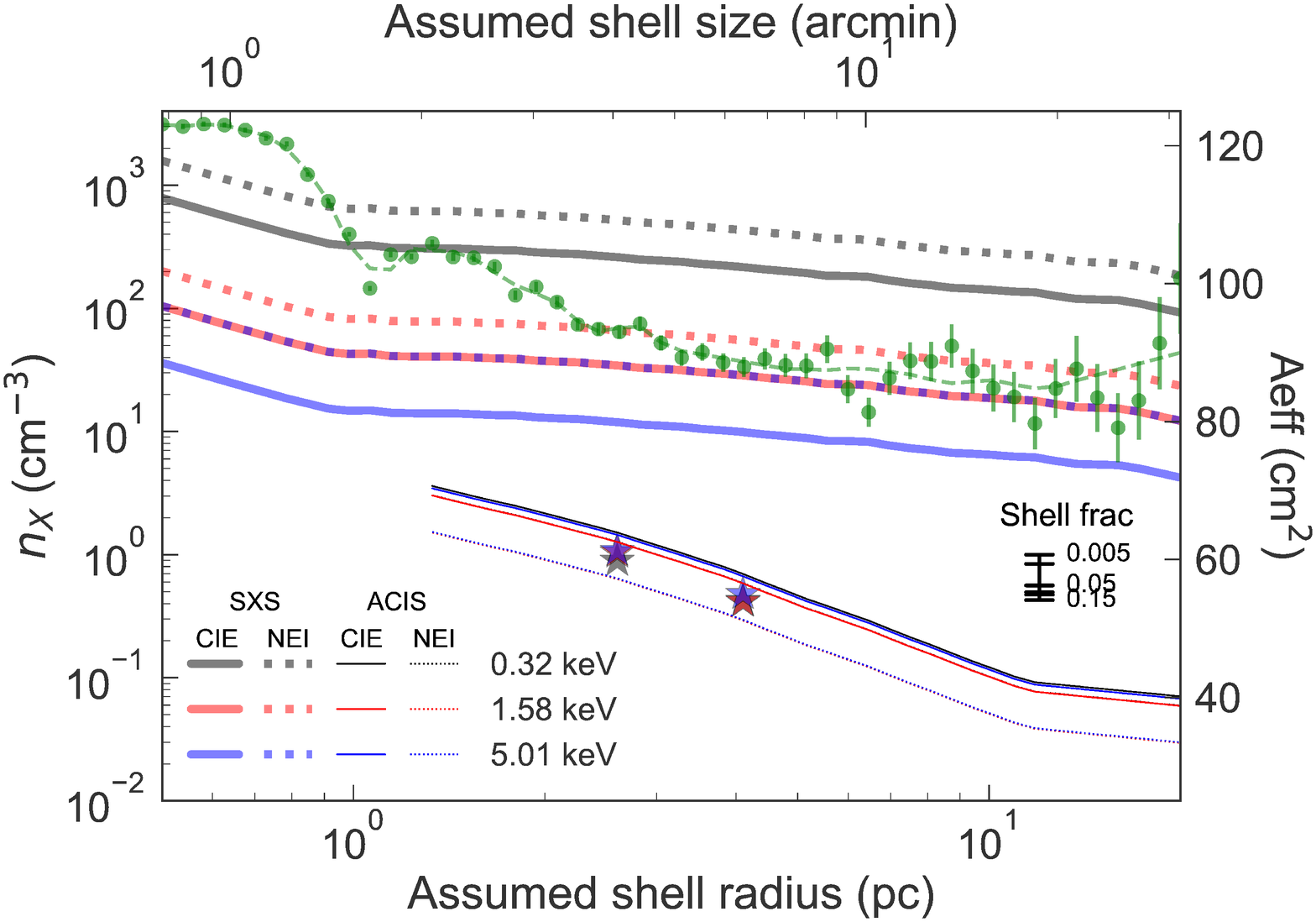}
 \end{center}
 \caption{Upper limits to the plasma density for several selected electron temperatures
 of a CIE (solid) and an NEI with $n_{\mathrm{e}}t=10^{10.5}$~s~cm$^{-3}$ (dotted)
 plasmas as a function of the assumed shell radius for the SXS (thick) and ACIS (thin;
 \cite{seward06}) when the shell fraction is $\Delta R/R=0.05$. The observed limits move
 vertically when the shell fraction is changed by the scaling shown in the figure. The
 effective area for the projected shell distribution is shown with green points with 
 statistical uncertainties by the ray-tracing simulations, which is smoothed (green
 dashes) by the \citet{savitzky64} method to use for the correction. The star marks are
 the expected limit with off-source pointing with the SXS at 2.6 and 4.1~pc for the CIE of
 different temperatures.}
\label{f08}
\end{figure}

We first use the upper limit of the plasma emission. The density is $n_{\mathrm{X}} =
\sqrt{Y/V_{\mathrm{obs}}}$, in which $V_{\mathrm{obs}}$ is the observed emitting
volume. Some selected cases are shown in figure~\ref{f08} (thick solid and dashed
curves). If the SXS square field of view with $\theta_{\mathrm{SXS}}=$3\farcm0 covers
the entire shell at $R<$~1\farcm3, $V_{\mathrm{obs}} \sim 4 \pi R^{2} \Delta R$. If the
field is entirely contained in the shell at $R>$~2\farcm1, $V_{\mathrm{obs}}$ should be
replaced with $\sim (D\theta_{\mathrm{SXS}})^{2}\Delta R$, in which $D$ is the distance
to the source. These approximations at the two ends make a smooth transition.

Here, we made a correction for the reduced effective area for the extended structure of
the shell. As $R$ increases within the SXS field of view, the effective area averaged
over the view decreases as more photons are close to the field edges. This effect is
small in the case of the Crab because the central pixels suffer dead time due to the high
count rate (figure~\ref{f10}). In fact, a slightly extended structure up to
$R\sim$1\farcm2 has a larger effective area than a point-like distribution. As $R$
increase beyond the field, the emission within the field becomes closer to a flat
distribution, and the reduction of the effective area levels off (figure~\ref{f08};
green data and dashed curve).

\begin{figure}
 \begin{center}
  \includegraphics[width=0.5\textwidth]{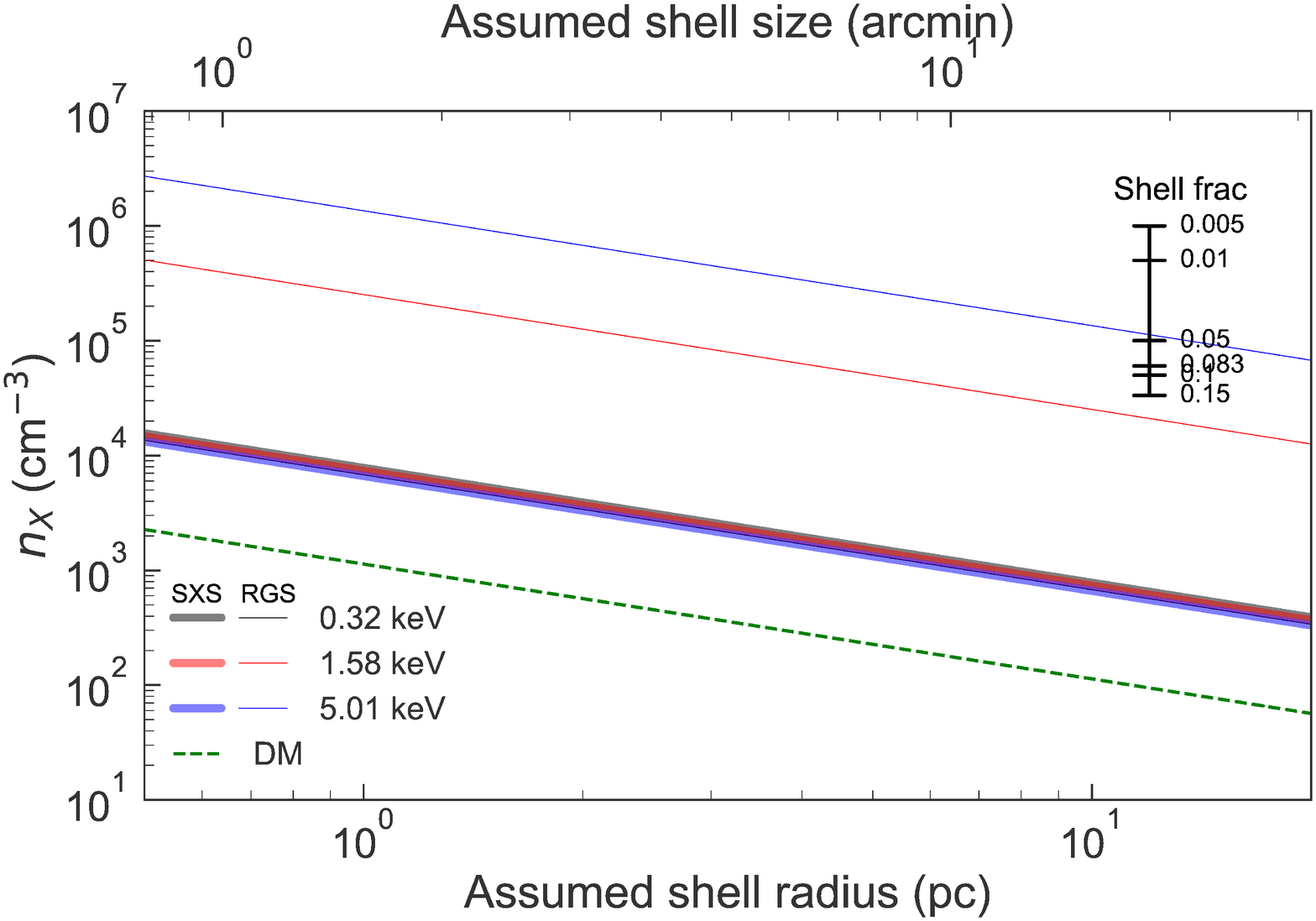}
 \end{center}
 \caption{Upper limits to the plasma density for several selected electron temperatures
 of a CIE plasma as a function of the assumed shell radius for the SXS (thick) and RGS
 (thin; \cite{kaastra09}) when the shell fraction is $\Delta R/R=0.05$. The observed
 limits move vertically when the shell fraction is changed by the scaling shown in the
 figure. Also shown is the upper limit by a radio dispersion measure (DM) of the Crab
 pulsar \citep{lundgren95}.}
\label{f07}
\end{figure}

Next, we convert the upper limits by the extinction column to the density with
$n_{\mathrm{X}} = N_{\mathrm{H}}^{\mathrm{(hot)}}/\Delta R$, which is shown in
figure~\ref{f07} (thick lines). We assume that the absorption feature is superposed on a
point-like continuum source, thus no correction was made for the extended structure.

\subsection{Results with other techniques}\label{s4-2}
We compare the results with the previous work using three different techniques. First,
\citet{seward06} used the Advanced CCD Imaging Spectrometer (ACIS; \cite{garmire03})
onboard the Chandra X-ray Observatory \citep{weisskopf02} with an unprecedented imaging
resolution, and derived the upper limit of the thermal emission assuming that it would
be detectable if it has a 0.1 times surface brightness of the observed halo emission
attributable to the dust scattering. We re-evaluated their raw data (their figure 5)
under the same assumptions with SXS (figure~\ref{f08}; thin solid and dashed curves). No
ACIS limit was obtained below $R \sim 2$\arcmin\ due to the extreme brightness of the
PWN. Beyond $R \sim 18$\arcmin, at which there is no ACIS measurement, we used the upper
limit at 18\arcmin. For the ACIS limits, a more stringent limit is obtained for the NEI
case with a low ionization age (10$^{10.5}$~s~cm$^{-3}$) than the CIE case with the same
temperature. This is because the Fe L series lines are enhanced for such NEI plasmas and
the ACIS is sensitive also at $<$2~keV unlike the SXS with the gate valve closed.

Second, \citet{kaastra09} presented the Crab spectrum using the Reflecting Grating
Spectrometer (RGS; \cite{denherder01}) onboard the XMM-Newton Observatory
\citep{jansen01} observatory. Upon the non-thermal emission of the PWN, they reported a
detection of the absorption feature by the O-He$\alpha$ and O-Ly$\alpha$ lines
respectively at 0.58 and 0.65~keV with a similar equivalent width of $\sim$0.2~eV
assuming that the lines are narrow. The former was also confirmed in the Chandra Low
Energy Transmission Grating data. However, these absorption lines are often seen in the
spectra of Galactic X-ray binaries (e.g., \cite{yao06a}), which is attributed to the hot
gas in the interstellar medium with a temperature of a few MK. Adopting the value by
\citet{sakai14}, the expected column density by such a gas to the Crab is
$\sim$8$\times$10$^{18}$~cm$^{-2}$, which is non-negligible. We therefore consider that
the values measured with RGS are an upper limit for the plasma around the Crab. Using
the same assumptions with SXS, we re-evaluated the RGS limit (thin lines in
figure~\ref{f07}).

Third, the dispersion measure from the Crab pulsar reflects the column density of
ionized gas along the line of sight. This includes not only the undetected thermal
plasma around the Crab but also the hot and warm interstellar gas. \citet{lundgren95}
derived a measure 1.8$\times$10$^{20}$~cm$^{-2}$, which converts to another density
limit (dashed line in figure~\ref{f07}).

\smallskip

\begin{figure}
 \begin{center}
  \includegraphics[width=0.5\textwidth]{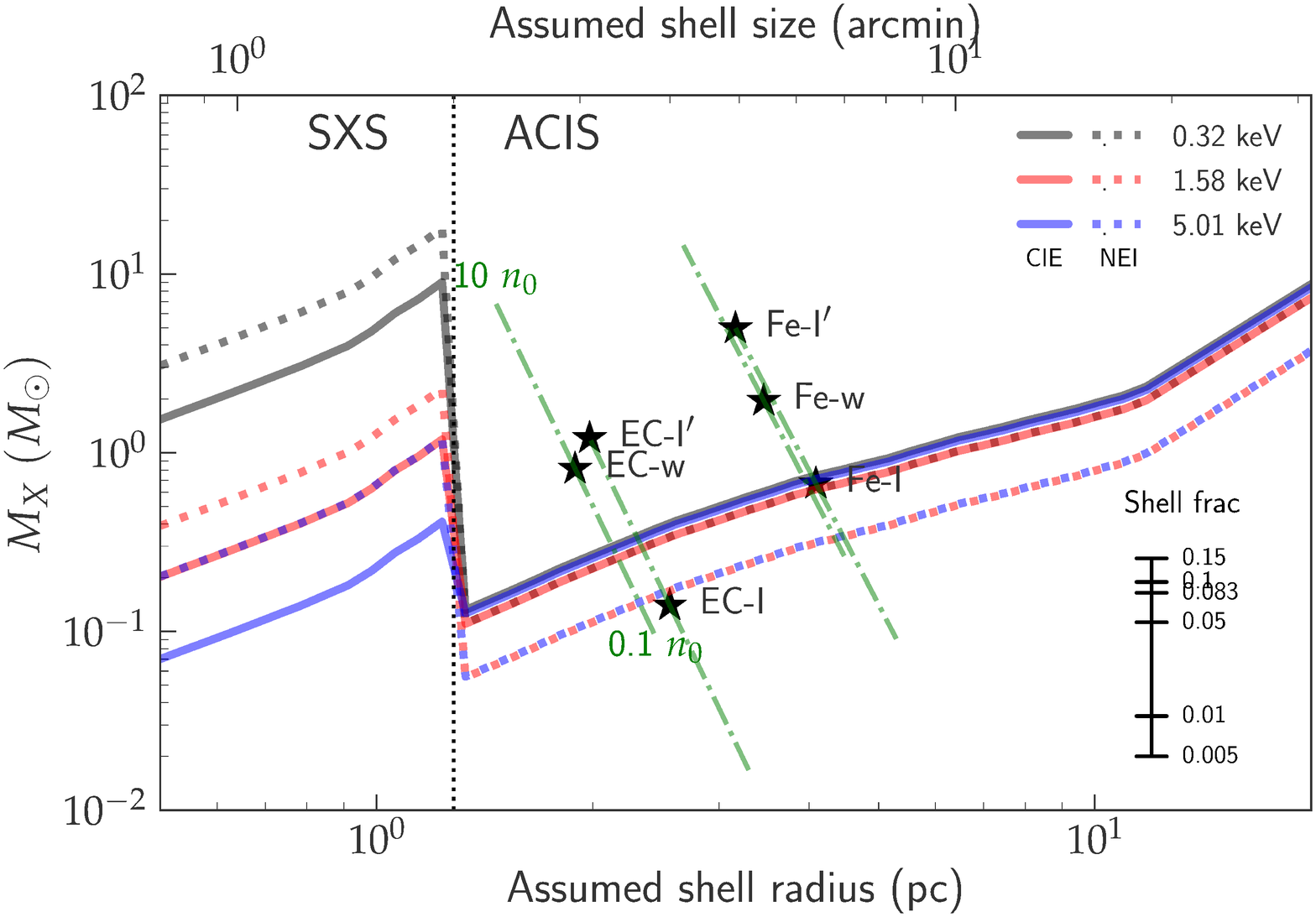}
 \end{center}
 \caption{Upper limit of the total plasma mass when the shell has a size $R$ for several
 electron temperatures of the CIE (solid) and NEI with
 $n_{\mathrm{e}}t=10^{10.5}$~s~cm$^{-3}$ (dotted) plasmas. $\Delta R/R=0.05$ is
 assumed. The observed limits move vertically when the shell fraction is changed by the
 scaling shown in the figure. The position of ($R_{\mathrm{CD}}$,
 $\overline{M_{\mathrm{X}}}$) is shown for the models in table~\ref{t03} with the stars,
 and their direction of change when $n_{0}$ is changed by a factor of 10 or 0.1 (dotted-and-dashed
 green lines from the stars).}
\label{f11}
\end{figure}

We now have the upper limit on $n_{\mathrm{X}}$ for several sets of $R$, $\Delta R$, and
$T$ by assembling the lowest values among various methods (re)-evaluated under the same
assumptions. We convert the limit to that of the total X-ray emitting mass
$M_{X}=n_{\mathrm{X}}m_{\mathrm{p}}V_{\mathrm{tot}}$, where $m_{\mathrm{p}}$ is the
proton mass and $V_{\mathrm{tot}}$ is the total emitting volume for an assumed shell
size and fraction. The resultant limit is shown in Figure~\ref{f11}. The most stringent
limit is given by the emission search either by ACIS or SXS. The SXS result complements
the ACIS result at $R<1.3$~pc, and the two give an upper limit of $\sim$1 $M_{\odot}$
for the X-ray emitting plasma at any shell radius. The exception is for the low plasma
temperature below $\sim$1~keV, for which the SXS with the closed gate valve yields a
less constraining limit.

\subsection{HD calculation}\label{s4-3}
We performed a HD calculation to verify that the searched parameter ranges
(table~\ref{t02}) are reasonable and to confirm if there are any SN models consistent
with the observed limit. We used the \textit{CR-hydro-NEI} code (\cite{lee14} and
references therein), which calculates time-dependent, non-equilibrium plasma in one
dimension. At the forward shock, the kinetic energy is thermalized independently for
each species, thus the temperature is proportional to the mass of the species. The
plasma is then thermally relaxed by the Coulomb interaction. No collissionless shocks
are included. Energy loss by radiation is included, while that by cosmic rays is
omitted.

We considered two SN explosion models under two circumstellar environments
(table~\ref{t03}) as representatives. The former two are (a) the Fe-core collapse SN by
a red super-giant progenitor with the initial explosion energy $E_0=1.21 \times
10^{51}$~erg and the ejecta mass $M_{\mathrm{ej}}=12.1$~$M_{\odot}$ \citep{patnaude15},
and (b) the electron capture (EC) SN by a super AGB progenitor with $E_0=0.15 \times
10^{51}$~erg and $M_{\mathrm{ej}}=4.36$~$M_{\odot}$ \citep{moriya14}. The latter two are
(1) the uniform density $n_0=0.1$~cm$^{-3}$ and (2) the density profile by the
progenitor wind: $n_{0}(r) = \dot{M}_{\mathrm{wind}} / (4 \pi v_{\mathrm{wind}}
m_{\mathrm{p}}r^{2})$, in which the mass loss rate $\dot{M}_{\mathrm{wind}} = 1 \times
10^{-5} M_{\odot}$~yr$^{-1}$ and the wind velocity $v_{\mathrm{wind}} = 20$~km~s$^{-1}$
\citep{moriya14}. In the wind density parameter \citep{chugai94}, $w =
\dot{M}_{\mathrm{wind}}/v_{\mathrm{wind}} = 3.2 \times 10 ^{14}$~g~cm$^{-1}$.

The 2$\times$2 models are labeled as (a-1) Fe-I, (a-2) Fe-w, (b-1)
EC-I, and (b-2) EC-w. For Fe-I and EC-I models, we also calculated an elevated ISM
density of $n_0=1.0$~cm$^{-3}$ (respectively labeled as Fe-I$^{\prime}$ and
EC-I$^{\prime}$). For all these models, we assumed the power ($n_{\mathrm{ej}}$) of the
unshocked ejecta density as a function of velocity to be 9 \citep{fransson96}. Only for
the model EC-w, we calculated with $n_{\mathrm{ej}}=7$ to see the effect of this
parameter (labeled as EC-w$^{\prime\prime}$).

\begin{table*}[ht]
  \caption{Result of HD calculation.}
  \label{t03}
 \begin{center}
  \begin{tabular}{lrrrrrrr}
   \hline
   Label                                    & Fe-I & Fe-I$^{\prime}$ & Fe-w & EC-I & EC-I$^{\prime}$ & EC-w & EC-w$^{\prime\prime}$ \\
   \hline
   \multicolumn{8}{l}{(SN setup)}\\
   SN explosion                             & Fe   & Fe   & Fe   & EC   & EC   & EC   & EC \\
   $E_{0}$ (10$^{51}$ erg)                  & 1.21 & 1.21 & 1.21 & 0.15 & 0.15 & 0.15 & 0.15 \\
   $M_{\mathrm{ej}}$ ($M_{\mathrm{\odot}}$) & 12.1 & 12.1 & 12.1 & 4.36 & 4.36 & 4.36 & 4.36 \\
   $n_{\mathrm{ej}}$                        & 9    & 9    & 9    & 9    & 9    & 9    & 7 \\
   Environment                              & ISM  & ISM  & wind & ISM  & ISM  & wind & wind \\
   $n_0$ (cm$^{-3}$)                        & 0.1  & 1.0  & ---  & 0.1  & 1.0  & ---  & --- \\
   $w=\dot{M}_{\mathrm{wind}}/v_{\mathrm{wind}}$ (10$^{14}$ g~cm$^{-1}$)   & ---  & ---  & 3.2 & ---  & ---  & 3.2 & 3.2 \\
   \hline
   \multicolumn{8}{l}{(SNR outcome)}\\
   $R_{\mathrm{FS}}$ (pc)                               & 4.6  & 3.6 & 4.3  & 2.9  & 2.2 & 2.3 & 2.6 \\
   $R_{\mathrm{CD}}$ (pc)                               & 4.1  & 3.2 & 3.5  & 2.6  & 2.0 & 1.9 & 2.0 \\
   $R_{\mathrm{RS}}$ (pc)                               & 3.8  & 2.9 & 3.3  & 2.4  & 1.8 & 1.8 & 1.9 \\
   $v_{\mathrm{FS}}$ (10$^{3}$ km~s$^{-1}$)             & 3.1  & 2.4 & 3.7  & 2.0  & 1.5 & 2.0 & 2.1 \\
   $v_{\mathrm{RS}}$\footnotemark[$*$] (10$^{3}$ km~s$^{-1}$)  & 1.4 & 1.2  & 0.51 & 0.88& 0.68 & 0.29 & 0.39 \\
   $M_{\mathrm{CD-FS}}$ ($M_{\odot}$)                   & 1.4  & 6.6 & 2.0  & 0.35 & 1.6 & 1.1 & 1.2 \\
   $M_{\mathrm{RS-CD}}$ ($M_{\odot}$)                   & 1.8  & 7.0 & 4.1  & 0.42 & 2.2 & 2.2 & 1.3 \\
   \multicolumn{8}{c}{------ derived values ------}\\
   $\frac{R_{\mathrm{CS}}-R_{\mathrm{RS}}}{R_{\mathrm{CD}}}$  & 0.07 & 0.07 & 0.04 & 0.06 & 0.09 & 0.04 & 0.07 \\
   $\frac{3}{16} \mu m_{\mathrm{p}}v_{\mathrm{FS}}^2$ (keV)   & 9.4  & 5.7  & 13   & 3.8  & 2.2 & 4.0  & 4.1 \\
   $M_{\mathrm{unshocked}}$ ($M_{\odot}$)               & 10  & 5.1  & 8.0  & 3.9  & 2.2 & 2.2 & 3.0 \\
   \multicolumn{8}{c}{------ absorbed X-ray flux weighted average ------}\\
   $\overline{T_{\mathrm{e}}}$ (keV)                    & 1.0  & 1.6 & 0.51 & 0.71 & 0.95 & 0.74 & 0.51 \\
   $\overline{T_{\mathrm{Fe}}}$ (keV)                   & 130  & 26  & 50   & 57   & 4.0  & 62  & 90 \\
   $\overline{n_{\mathrm{e}}t}$ (10$^{11}$~cm~s$^{-1}$) & 0.21 & 1.5 & 9.9  & 0.22 & 1.59 & 11.8 & 10.2 \\
   $\overline{M_{\mathrm{X}}}$ ($M_{\odot}$)            & 0.67 & 5.0 & 2.0  & 0.14 & 1.2  & 0.81 & 1.1 \\
   \hline
   \multicolumn{8}{l}{\parbox{70mm}{
   \footnotesize
   \par \noindent
   \footnotemark[$*$] Velocity with respect to the ejecta.\\
   }}
  \end{tabular}
 \end{center}
\end{table*}

Table~\ref{t03} summarizes the SN setup stated above and the SNR outcome at an age of
962~yr, which includes the radius of the forward shock (FS), contact discontinuity (CD),
and reverse shock (RS) ($R_{\mathrm{FS}}$, $R_{\mathrm{CD}}$, and $R_{\mathrm{RS}}$),
the velocity of the forward and reverse shocks ($v_{\mathrm{FS}}$ and
$v_{\mathrm{RS}}$), the mass between CD and FS ($M_{\mathrm{CD-FS}}$) and that between
RS and CD ($M_{\mathrm{RS-CD}}$). The two masses represent the shocked ISM and ejecta,
respectively. The radius is close to the observed size of the optical photo-ionized
nebula, and the radii and velocities match reasonably well with analytical approaches
\citep{chevalier82,truelove99} within 10\%, which validates our calculation. The RS
radius is larger than the X-ray emitting synchrotron nebula, which justifies that our
calculation does not include the interaction with it.

From these, we calculated $(R_{\mathrm{RS}}-R_{\mathrm{CD}})/R_{\mathrm{CD}}$ as a proxy
for the shell fraction, $3 \mu m_{\mathrm{p}}v_{\mathrm{FS}}^2/16$ as a proxy for the
electron temperature after Coulomb relaxation, in which $\mu=0.5$ is the mean molecular
weight, and the unshocked ejecta mass $M_{\mathrm{unshocked}} = M_{\mathrm{ej}} -
M_{\mathrm{RS-CD}}$. We also derived the average of the electron and Fe temperatures
($\overline{T_{\mathrm{e}}}$ and $\overline{T_{\mathrm{Fe}}}$) and the ionization age
($\overline{n_{\mathrm{e}}t}$) weighted over the absorbed X-ray flux. The X-ray emitting
mass ($\overline{M_{\mathrm{X}}}$) was estimated by integrating the mass with a
temperature in excess of $\overline{T_{\mathrm{e}}}$.

The searched ranges of all parameters (table~\ref{t02}) encompass the HD result for
all models. The electron temperature is expected between $3 \mu
m_{\mathrm{p}}v_{\mathrm{FS}}^2/16$ and $\overline{T_{\mathrm{e}}}$; the former is the
highest for thermalizing all the kinetic energy instantaneously, while the latter is the
lowest for starting the Coulomb relaxation without collision-less heating.  The averaged
Fe temperature $\overline{T_{\mathrm{Fe}}}$ is sufficiently low to consider that the
line is relatively narrow; the thermal broadening by this is 32~eV at 6.7~keV for
$\overline{T_{\mathrm{Fe}}}$=130~keV. The ionization age ($n_{\mathrm{e}}t$) ranges over
two orders from 10$^{10}$ to 10$^{12}$ cm$^{-3}$~s$^{-1}$ depending on the pre-explosion
environment, where the wind density cases result in higher values than the ISM density
cases.

\subsection{Comparison with observed limits}\label{s4-4}
Finally, we compare the HD results with the observation in figure~\ref{f11}. For the
radius and the X-ray plasma mass, we plotted ($R_{\mathrm{CD}}$,
$\overline{M_{\mathrm{X}}}$) in table~\ref{t03}. The shell size by the models
($R_{\mathrm{CD}}$) is larger than 1.3~pc, where we have a stringent limit on
$M_{\mathrm{X}}$ with the observations. The HD results depend on the choice of the
parameters in the SN setup ($E_{0}$, $M_{\mathrm{ej}}$, $n_{0}$ or $w$, and
$n_{\mathrm{ej}}$; table~\ref{t03}). We can estimate in which direction the model points
move in the plot when these parameters are changed.

First, the two parameters $E_{0}$ and $M_{\mathrm{ej}}$ are known to be correlated in
type II SNe. Our two SN models are in line with the relation by
\citet{pejcha15}. Therefore, the model points move roughly in the direction of the lines
connecting the EC-I and Fe-I models, or the EC-w and Fe-w models. For a fixed explosion
energy of 1.21$\times$10$^{51}$~erg for our Fe model, a plausible range of
$M_{\mathrm{ej}}$ is 12--32~$M_{\odot}$ \citep{pejcha15}, thus our model is close to the
lower bound. Second, for $n_{0}$, the points move in parallel with the lines connecting
Fe-I and Fe-I$^{\prime}$ or EC-I and EC-I$^{\prime}$. This should be the same for $w$ in
the wind environment case. Third, for $n_{\mathrm{ej}}$, there is little difference
between the result of the model Fe-w and Fe-w$^{\prime\prime}$, so we consider that this
parameter does not affect the result very much. In terms of the comparison with the
observation limit, $n_{0}$ or $w$ is the most important factor.

Although the small observed mass of the Crab is argued to rule out an Fe core collapse
SN for its origin \citep{seward06}, we consider that this does not simply hold. Our
models illustrate that such a small mass can be reproduced if an Fe core collapse SN
explosion takes place in a sufficiently low density environment with the ISM density
$n_{0} \lesssim$ 0.03~cm$^{-3}$ (Fe-I) or the wind density parameter $w \lesssim
10^{14}$~g~cm$^{-1}$ (Fe-w). In such a case, a large fraction of the ejecta mass is
unshocked (table~\ref{t03}) and escapes from detection. Some of the unshocked ejecta may
be visible when they are photo-ionized by the emission from the PWN to a
$\approx$10$^{3}$~K gas \citep{fesen97} or a $\approx$10$^{4}$~K gas
\citep{sollerman00}.

We argue that both the Fe and EC models still hold to be compatible with the observed
mass limits. In either case, it is strongly preferred that the pre-explosion environment
is low in density; i.e., $n_{0} \lesssim 0.1$~cm$^{-3}$ (EC-I) or $\lesssim
0.03$~cm$^{-3}$ (Fe-I) for the ISM environment or $w \lesssim 10^{14}$~g~cm$^{-1}$ for
the wind environment (both Fe-w and EC-w). For the latter, a large $w$ value (e.g., $6
\times 10^{18}$~g~cm$^{-1}$; \cite{smith13}), which is an idea to explain the initial
brightness of SN\,1054, is not favored. In fact, such a low density environment is
suggested by observations. At the position of the Crab, which is off-plane in the
anti-Galactic center direction, the ISM density is $\sim$0.3~cm$^{-3}$ by a Galactic
model \citep{ferriere98}. \citet{wallace99} further claimed the presence of a bubble
around the Crab based on an H I mapping with a density lower than the surroundings. Our
result suggests that SN\,1054 took place in such a low $n_{0}$ environment and the wind
environment by its progenitor of a low wind density value.

\section{Conclusion}\label{s5}
We utilized the SXS calibration data of the Crab nebula in 2--12~keV to set an upper
limit to the thermal plasma density by spectroscopically searching for emission or
absorption features in the Crab spectrum. No significant emission or absorption features
were found in both the plasma and the blind searches.

Along with the data in the literature, we evaluated the result under the same
assumptions to derive the X-ray plasma mass limit to be $\lesssim 1 M_{\odot}$ for a
wide range of assumed shell radii ($R$) and plasma temperatures ($T$). The SXS sets
a new limit in $R<$~1.3~pc for $T>1$~keV. We also performed HD simulations of the
Crab SNR for two SN explosion models under two pre-explosion environments. Both SN models
are compatible with the observed limits when the pre-explosion environment has a low
density of $\lesssim$ 0.03~cm$^{-3}$ (Fe model) or $\lesssim$ 0.1~cm$^{-3}$ (EC model)
for the uniform density, or $\lesssim$ 10$^{14}$~g~cm$^{-1}$ ($\dot{M}_{\mathrm{wind}}
\lesssim 3 \times 10^{-6} M_{\odot}$~yr$^{-1}$ for $v_{\mathrm{wind}}=$~20~km~s$^{-1}$)
for the wind density parameter in the wind environment.

A low energy explosion is favored based on the abundance, initial light curve, and
nebular size studies \citep{macalpine08,moriya14,yang15}. We believe that a
positive detection of thermal plasma, in particular with lines, is key to
distinguishing the Fe and EC models. It is worth noting that the observed limit is close
to the model predictions. We now know the high potential of a spectroscopic search with
the SXS, and may expect a detection of the thermal feature by placing the SXS field
center at several offset positions. With a 10~ks snapshot at four different positions at
the radius of EC-I and Fe-I models (respectively 2.6 and 4.1~pc), an upper limit lower
than that with ACIS by a factor of a few is expected (figure~\ref{f08}).

This was exactly what was planned next. If it were not for the loss of the spacecraft
estimated to have happened at 1:42 UT on 2016 March 26, a series of the offset Crab
observations should have started 8 hours later for calibration purposes, which should
have been followed by the gate valve open to allow access down to 0.1~keV. The 8 hours
now turned to be many years, but we should be back as early as possible.

\begin{trueauthors}
 M. Tsujimoto led this study in data analysis and writing drafts. He also contributed to
 the SXS hardware design, fabrication, integration and tests, launch campaign, in-orbit
 operation, and calibration. S.-H. Lee performed the hydro calculations and its
 interpretations for this paper. K. Mori and H. Yamaguchi contributed to discussion on
 SNRs. They also made hardware and software contributions to the Hitomi
 satellite. N. Tominaga and T. J. Moriya gave critical comments on SNe. T. Sato worked
 for the telescope response for the data analysis and calibration. C. de Vries led the
 filter wheel of the SXS, which gave the only pixel-to-pixel gain reference of this
 spectrometer in the orbit. R. Iizuka contributed to the testing and calibration of the
 telescope, and the operation of the SXS. A. R. Foster and T. Kallman helped with the
 plasma models. M. Ishida, R. F. Mushotzky, A. Bamba, R. Petre, B. J. Williams,
 S. Safi-Harb, A. C. Fabian, C. Pinto, L. C. Gallo, E. M. Cackett, J. Kaastra, M. Ozaki,
 J. P. Hughes, and D. McCammon improved the draft.
\end{trueauthors}

\begin{ack}
 We appreciate all people contributed to the SXS, which made this work possible. We also
 thank Toru Misawa in Shinshu University for discussing the C IV feature.

 We thank the support from the JSPS Core-to-Core Program.
We acknowledge all the JAXA members who have contributed to the ASTRO-H (Hitomi)
project.
All U.S. members gratefully acknowledge support through the NASA Science Mission
Directorate. Stanford and SLAC members acknowledge support via DoE contract to SLAC
National Accelerator Laboratory DE-AC3-76SF00515. Part of this work was performed under
the auspices of the U.S. DoE by LLNL under Contract DE-AC52-07NA27344.
Support from the European Space Agency is gratefully acknowledged.
French members acknowledge support from CNES, the Centre National d'\'{E}tudes Spatiales.
SRON is supported by NWO, the Netherlands Organization for Scientific Research.  Swiss
team acknowledges support of the Swiss Secretariat for Education, Research and
Innovation (SERI).
The Canadian Space Agency is acknowledged for the support of Canadian members.  
We acknowledge support from JSPS/MEXT KAKENHI grant numbers 15H00773, 15H00785,
15H02090, 15H03639, 15H05438, 15K05107, 15K17610, 15K17657, 16H00949, 16H06342,
16K05295, 16K05300, 16K13787, 16K17672, 16K17673, 21659292, 23340055, 23340071,
23540280, 24105007, 24540232, 25105516, 25109004, 25247028, 25287042, 25400236,
25800119, 26109506, 26220703, 26400228, 26610047, 26800102, JP15H02070, JP15H03641,
JP15H03642, JP15H03642, JP15H06896, JP16H03983, JP16K05296, JP16K05309, JP16K17667, and
16K05296.
The following NASA grants are acknowledged: NNX15AC76G, NNX15AE16G, NNX15AK71G,
NNX15AU54G, NNX15AW94G, and NNG15PP48P to Eureka Scientific.
H. Akamatsu acknowledges support of NWO via Veni grant.  
C. Done acknowledges STFC funding under grant ST/L00075X/1.  
A. Fabian and C. Pinto acknowledge ERC Advanced Grant 340442.
P. Gandhi acknowledges JAXA International Top Young Fellowship and UK Science and
Technology Funding Council (STFC) grant ST/J003697/2. 
Y. Ichinohe, K. Nobukawa, H. Seta, and T. Sato are supported by the Research Fellow of JSPS for Young
Scientists.
N. Kawai is supported by the Grant-in-Aid for Scientific Research on Innovative Areas
``New Developments in Astrophysics Through Multi-Messenger Observations of Gravitational
Wave Sources''.
S. Kitamoto is partially supported by the MEXT Supported Program for the Strategic
Research Foundation at Private Universities, 2014-2018.
B. McNamara and S. Safi-Harb acknowledge support from NSERC.
T. Dotani, T. Takahashi, T. Tamagawa, M. Tsujimoto and Y. Uchiyama acknowledge support
from the Grant-in-Aid for Scientific Research on Innovative Areas ``Nuclear Matter in
Neutron Stars Investigated by Experiments and Astronomical Observations''.
N. Werner is supported by the Lend\"ulet LP2016-11 grant from the Hungarian Academy of
Sciences.
D. Wilkins is supported by NASA through Einstein Fellowship grant number PF6-170160,
awarded by the Chandra X-ray Center, operated by the Smithsonian Astrophysical
Observatory for NASA under contract NAS8-03060.

We thank contributions by many companies, including in particular, NEC, Mitsubishi Heavy
Industries, Sumitomo Heavy Industries, and Japan Aviation Electronics Industry. Finally,
we acknowledge strong support from the following engineers.  JAXA/ISAS: Chris Baluta,
Nobutaka Bando, Atsushi Harayama, Kazuyuki Hirose, Kosei Ishimura, Naoko Iwata, Taro
Kawano, Shigeo Kawasaki, Kenji Minesugi, Chikara Natsukari, Hiroyuki Ogawa, Mina Ogawa,
Masayuki Ohta, Tsuyoshi Okazaki, Shin-ichiro Sakai, Yasuko Shibano, Maki Shida, Takanobu
Shimada, Atsushi Wada, Takahiro Yamada; JAXA/TKSC: Atsushi Okamoto, Yoichi Sato, Keisuke
Shinozaki, Hiroyuki Sugita; Chubu U: Yoshiharu Namba; Ehime U: Keiji Ogi; Kochi U of
Technology: Tatsuro Kosaka; Miyazaki U: Yusuke Nishioka; Nagoya U: Housei Nagano;
NASA/GSFC: Thomas Bialas, Kevin Boyce, Edgar Canavan, Michael DiPirro, Mark Kimball,
Candace Masters, Daniel Mcguinness, Joseph Miko, Theodore Muench, James Pontius, Peter
Shirron, Cynthia Simmons, Gary Sneiderman, Tomomi Watanabe; ADNET Systems: Michael
Witthoeft, Kristin Rutkowski, Robert S. Hill, Joseph Eggen; Wyle Information Systems:
Andrew Sargent, Michael Dutka; Noqsi Aerospace Ltd: John Doty; Stanford U/KIPAC: Makoto
Asai, Kirk Gilmore; ESA (Netherlands): Chris Jewell; SRON: Daniel Haas, Martin Frericks,
Philippe Laubert, Paul Lowes; U of Geneva: Philipp Azzarello; CSA: Alex Koujelev, Franco
Moroso.

\end{ack}

\bibliographystyle{aa}
\bibliography{ms}

\end{document}